\documentclass{IOS-Book-Article}

\usepackage{mathptmx}
\usepackage{soul}\setuldepth{article}

\usepackage{graphicx}

\usepackage{balance}
\usepackage{threeparttable}
\usepackage{graphicx}
\usepackage{multirow}
\usepackage{booktabs}
\usepackage{tabularx}
\usepackage{array,makecell,booktabs}
\newcolumntype{M}[1]{>{\centering\arraybackslash}m{#1}}

\def\hb{\hbox to 11.5 cm{}}

\begin{document}

\pagestyle{headings}
\def\thepage{}

\begin{frontmatter}              

\title{Connected Vehicle Platforms for Dynamic Insurance}

\markboth{}{August 2022\hb}

\author[A]{\fnms{Christian} \snm{Colot}%
\thanks{Corresponding Author: Christian Colot; E-mail:
christian.colot@uni.lu}},
\author[A]{\fnms{François} \snm{Robinet}}
and
\author[B]{\fnms{Geoffrey} \snm{Nichils}}
\author[A]{\fnms{Raphaël} \snm{Frank}}

\runningauthor{Colot et al.}
\address[A]{SnT - Interdisciplinary Centre for Security, Reliability and Trust, University of Luxembourg, Avenue John F. Kennedy 29, 1855 Luxembourg Kirchberg, Grand Duchy of Luxembourg}
\address[B]{Foyer Assurances Luxembourg, rue Léon Laval 12, 3372 Leudelange, Grand Duchy of Luxembourg}

\begin{abstract}
Following a regulatory change in Europe which mandates that car manufacturers include an “eCall” system in new vehicles, many car manufacturers are adding additional services on top, so that more and more cars become connected vehicles and act like IoT sensors. In the following study, we analyse the maturity level of this new technology to build insurance products that would take vehicle usage into account. For this, the connectivity of recent cars a-priori eligible has been first tested. Then, an ad-hoc platform has been designed to collect driving data. In particular, 4 cars have been connected to this platform for periods of over one month. Our results highlight that, while this technological innovation appears very promising in the future, the pricing, the lack of uniformity of data collected and the enrollment process are currently three pain points that should be addressed to offer large-scale opportunities. In the meantime, this technology might still be used for high value use cases such as the insurance of luxurious cars.

\end{abstract}

\end{frontmatter}
\markboth{June 2021\hb}{June 2021\hb}

\section{Introduction}
In the past, cars have been largely conceived as mechanical vehicles. This has considerably changed in the last decades, as cars have been more and more equipped with electronic devices \cite{haberle2015}. Today, it is difficult to identify driving related information and tasks that cannot be covered by these devices, with the notable exception of autonomous driving which is still on its way \cite{gao2021}.  

Besides helping the driver on the road, these devices also trigger new opportunities for better decision-making related to various purposes. This is the case for car insurances \cite{baecke2017}. Before, insurance companies were restricted to basic information related to drivers and to their cars, such as the age of the driver and the horse power of the vehicle, to define the insurance premium \cite{denuit2007}. This pricing system is particularly unfair for good drivers belonging to a-priori risky groups such as young drivers. Furthermore, this system does not foster better driving practices on the road \cite{stigson2014}. 

There have been some attempts to mitigate these issues without technological support, such as insurances based on the reported number of kilometers travelled (also known as "Pay as You Drive") or based on accident history \cite{edlin1999}. However, these initiatives led to limited outcomes. Indeed, both of them are subject to fraud \cite{boucher2009}. In particular, accidents leading to small repair costs are typically not reported to avoid a premium increase \cite{boucher2009}. Additionally, since car accidents remain rare events, the absence of historical claims for a particular driver does not necessarily imply safe driving on his/her part\cite{denuit2019}.

In the meantime, new car insurance products have been proposed by deploying new electronic devices in the vehicle. One is a retrofitted tracking system \cite{huang2019}, commonly referred to as “Black Box” which records driving information from the car and transmits it to the insurance company on a regular basis. This system however several drawbacks. First, the users are reluctant to install such devices in their private cars, as they are usually perceived as invasive. Second, the cost of such systems is usually significant, hence reducing the margin for the insurance provider who often covers their cost. Other devices commonly used are OBD dongles \cite{ullah2020}. Like a Black Box, such a device is plugged into the vehicle and records some information of the car such as speed or consumption. This system is cheaper than a Black Box, but it might still be considered as invasive by the driver and it might also be subject to fraud (i.e. device disconnected). Furthermore, it also captures limited information of interest for car insurance as it has been developed to retrieve diagnostics on the state of the vehicle and not for insurance purposes. A third device is the smartphone \cite{castignani2015}. There are currently some third party apps running on smartphones which compute insurance premiums based on recorded information. This technology however suffers from two main drawbacks: risk of fraud (e.g. smartphone placed in another car or smartphone turned off when driving fast) and the low level of accuracy provided by sensors \cite{handel2014}. Overall, these electronic devices are not suitable for insurance products mainly because they are not natively embedded in the vehicle.  

In Europe, as of March 2018, a regulatory change has fostered the emergence of an electronic device connected by default \cite{eu2014}. More concretely, every new vehicle has to be equipped with the so called “eCall” system which allows to locate and communicate with a car if a crash is detected. This system is composed of a positioning system (e.g. GPS) and a communication module (e.g. 3G/4G/5G connectivity). Many car manufacturers are therefore adding additional services on top of eCall to generate a new revenue stream \cite{mcdonnell2021}. 

In this study, we investigate to which extent this embedded device might be leveraged for an \textit{Usage-Based Insurance} (UBI). In particular, we propose a platform to collect data and report insights based on tests performed with real cars. The rest of the paper is organised as follows. In Section~\ref{sec:related_work}, the related literature on UBI is discussed. Then, in Section~\ref{sec:platform}, a specific architecture to collect the data of interest is proposed. Next, in Section~\ref{sec:test}, the tests made and results presented. Afterwards, in Section~\ref{sec:discussion}, a discussion on the opportunity to use this new approach for insurance products is initiated. Finally, in Section \ref{sec:conclusion}, we conclude this paper and discuss future works.   

\section{Related Work on Usage-Based Insurance}\label{sec:related_work}

In the insurance industry, \textit{Usage-Based Insurance} (UBI) consists of taking into account the usage of the insured risk to define the corresponding insurance offer. In particular for car insurance policies \cite{bian2018}, there are two specific ways to address this usage: \textit{Pay As You drive} (PAYD) and \textit{Pay How You drive} (PHYD). While PAYD policies use merely basic usage information of the car to compute the premium such as the number of kilometers driven, PHYD additionally include information related to the driving style such as the number of kilometers over the legal limit or the frequency of hard brakes \cite{castignani2015}. 

More concretely, whatever the perspective chosen, usage data primarily permits to better assess the risk component of the premium. A significant number of scientific articles confirm the additional value of this information to better predict car accident \cite{baecke2017,huang2019,paefgen2013,guillen2019}. In particular, the total distance travelled plays an important role \cite{baecke2017,paefgen2013}. Basically, more distance means more exposure to the risk. However, this effect is decreasing over the distance as larger distances mean also more driving experience \cite{guillen2019}. Additional factors include driving style such as harsh braking \cite{castignani2015}, sudden acceleration \cite{huang2019} and contextual factors such as night driving \cite{baecke2017,huang2019}, driving in urban places \cite{baecke2017,huang2019}. Furthermore, some studies highlight additional benefits of using UBI insurance beyond a better risk estimation. These benefits include a more responsible driving style \cite{stigson2014,tselentis2016}, less pollution \cite{litman2007,stigson2014} and a better traffic flow \cite{litman2007,mcdonnell2021}.

For these reasons, this new way of conceiving car insurance products is attractive. This is especially the case for insurance companies and good drivers. However, UBI has not met success in many countries. This is mainly due to the following reasons: cost \cite{litman2007}, reliability of the system \cite{handel2014} and customer engagement \cite{litman2007,kaiser2018}. In the following study, we assess to which extent embedded devices might mitigate these drawbacks, so that UBI might enter in a new era.   

\section{PLATFORM}\label{sec:platform}

\subsection{Design Choice}

To design a platform collecting data of connected cars, i.e. cars including the “eCall” system, one has first to decide the depth of the flow managed internally. Basically, the connected car only communicates directly with its OEM (Original Equipment Manufacturer). Then, the OEM can communicate data to another organisation as long as the driver gives her/his consent. This organisation might be directly the company which will make use of the data i.e. in our case an insurance company or a third-party data aggregator which collects data from every OEM to transmit it to the insurance company.

The insurance company should consequently decide if it prefers to handle the data gathering process with every OEM separately or to do it with one third-party. Dealing directly with each OEM has the advantage of avoiding the cost of an additional party. However, this has the disadvantage of having to negotiate with each of them and to design a specific data gathering process for each OEM. This process is typically different from one OEM to another as there is no standardised way to design this process among OEMs (at least so far) neither on the data points provided nor on the technical specifications to retrieve it. Furthermore, OEMs are making changes on their process over time to include more data points. For these reasons, we decided in this study to set up our data gathering process with one single third-party. Most of the conclusions of the tests that will be presented in Section 4 are however also valid if one chooses to set up the data gathering process with every OEM separately.

\subsection{Description of the Platform}
   \begin{figure}[!t]
      \centering
      \includegraphics[scale=0.25]{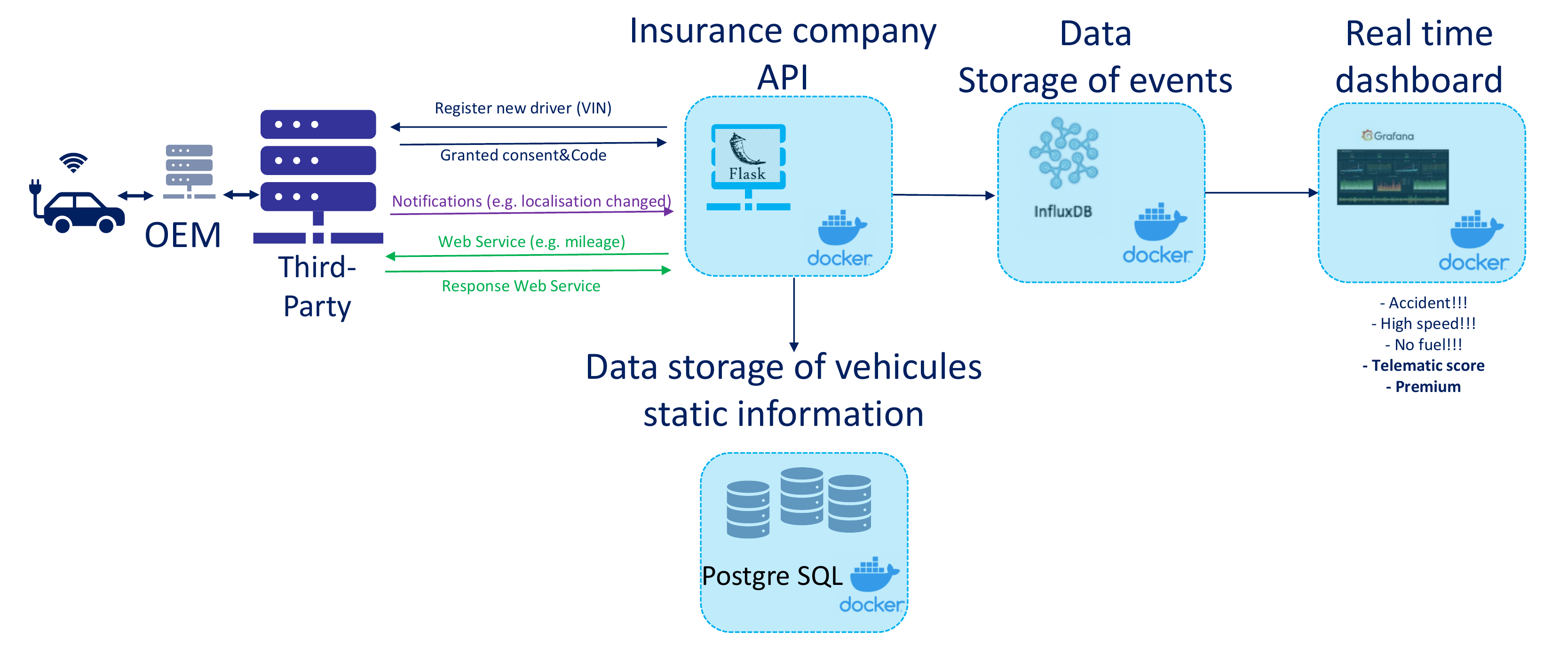}
      \caption{Architecture of the platform}
      \label{architecture_figure}
   \end{figure}
The architecture of the platform is displayed in figure \ref{architecture_figure}. The platform is a Flask server i.e. an API using a graphical interface. There are three types of flow exchanges between the platform and the connected car: (1) consent, (2) notification and (3) request.

The consent flow organises the request to ask the consent of the driver for collecting data of his/her car. This request is done on the platform by introducing the Vehicle Information Number (VIN) of the car and the email address of the driver. Then, an email is sent to the driver. Next, the driver logs in the platform of the third-party trough a hyperlink in the email. On this platform, whatever the OEM, the owner specifies the OEM of the car, reviews the information that would transmitted and gives consent in compliance with GDPR regulation \cite{eu,european2020}. Beside this, the remaining of consent flow is handled by the concerned OEM which ensures a.o. that the declared driver is the owner or the leaser of the car. This is typically performed by using a platform of the OEM where the driver needs to log in. More details by OEM are given in Section IV.

Once this step is completed, our platform receives a confirmation of the consent and technical information needed to retrieve information from the car by using an OAuth 2.0 protocol \cite{hardt2012}.

The notification flow is triggered by a specific event such as a change of localisation detected by the connected car. It sends a webhook to our platform through the OEM and the third-party to notify the occurence of this event. The list of events available depends on the OEM. See table \ref{tab:data-points} for the list of events for the tested OEMs. Within this list, it is also possible to select only some of them as it might have an influence on the price to get these data. This choice is made a-priori with the third-party and applies for all cars of the OEM for which a consent is granted.     

The request flow is dedicated to handle requests initiated by the insurance company to get data from a connected car. This is only possible for cars for which a consent is valid. Note that a request can be done in reaction to a notification received. For instance, if the platform receives the information that a car has changed of position, the request might ask the new GPS coordinates of the car. A request might also be done without any link with a specific event related to the car. For example, it is possible to ask every day the mileage of the car and hence computing the number of kilometers driven each day. See table \ref{tab:data-points} for the list of available requests for the tested OEMs.

Once new information is transmitted to the platform through one of these three processes, it has to be stored in a database. We have deployed two databases: one dedicated to store static information and one for dynamic information. The storage of static information is setup with a PostgreSQL DBMS. It includes information related to the driver and the car retrieved during the consent flow. The storage of dynamic information is done with InfluxDB. This is a time series DB which records details of information retrieved with the notification flow (if relevant) and request flow along with the time of the event. Note that our platform does not register the notification of location change as this notification triggers a request to get the new GPS coordinates of the vehicle which is more useful that the notification itself. 

\begin{figure}[t]\vspace*{4pt}
\centerline{\includegraphics[width=50mm,scale=0.5]{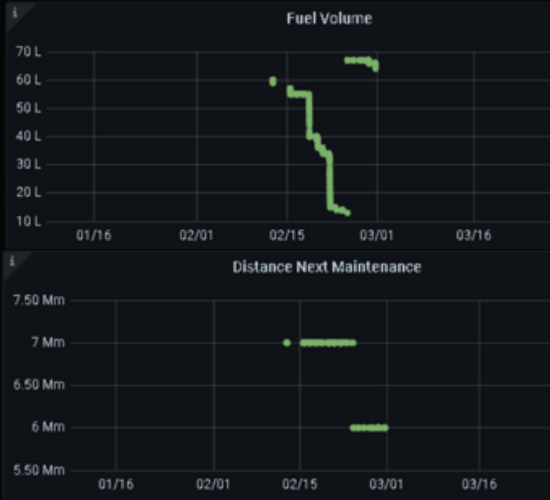}}
\caption{\label{fig:odometer}Evolution of fuel volume and distance to next maintenance for BMW 116d from 2022/02/15 to 2022/02/31}\vspace*{-6pt}
\end{figure}

The information stored in the databases can then be used in multiple ways. We have designed a realtime dashboard which displays the static and dynamic information of a specific vehicle. As an example, Figure \ref{fig:odometer} exhibits the evolution of the fuel volume and distance to next maintenance for one of the connected cars. The aim is to get more insights during the tests of the platform and to be able communicating more smoothly with people involved in the research project that do not have a technical background. Another usage is for insurance premium computation once tests are done. This part has not been integrated in the platform as it is out of scope of the study. We give however some details about what it could look like. Once enough historical data has been collected on a significant number of vehicles, this data along with data internally recorded by the insurance company can be used to assess the predicted number of accidents on a given period of time with a machine learning software such as TensorFlow. Then, based on this prediction, an insurance premium can be derived for each car insured trough actuarial computations.  

\section{TEST OF THE PLATFORM}\label{sec:test}

\subsection{Prerequisite: Check of Eligibility of Vehicle}

Before being able to connect a given car to the platform, it is necessary to ensure as much as possible that this vehicle is connectable. This check is done in two sequential steps: (1) requirement check and (2) VIN check. 

For the first step, OEMs give an overview of main requirements for this. These requirements typically include a list of models, year of production and the country where the car was bought. It may also include the participation to a specific fidelity program of the OEM. Cars that fulfill all these conditions are usually connectable. If only the condition regarding the year of production is unmet, the car might still be connectable but is is less likely to be the case. Other cars are not connectable. In this study, we investigated 19 cars that fulfilled the requirements of this first step.

For cars identified as potentially connectable during the first step, it is advised to further test the possibility of connectivity based on the step 2 of the eligibility check (i.e. the VIN check). Currently, this VIN check might be done either fully automatically trough an API or with a process including human intervention depending on the specific OEM (at least from the car manufacturer side). Note that when an manual intervention is needed, this step requires at least a couple of business days. Specifically for cars of the Stellantis group, beyond these 2 steps, there are additional verifications that should be done to ensure eligibility. These verifications are however integrated to the consent process that is discussed next. The statistics of the VIN checks for the 19 cars considered in this study are included in table \ref{tab:stats-eligibility}.

\begin{table}
\centering%
\caption{\label{tab:stats-eligibility}Statistics of eligibility by car}
\resizebox{0.45\textwidth}{!}{%
\begin{tabular}{c M{0.12\textwidth} M{0.12\textwidth}}
    \toprule%
        {\centering\bfseries OEM} & {\centering\bfseries Cars satisfying\\requirements} & {\centering\bfseries Cars satisfying\\VIN check}\\
    \midrule
        Alfa Romeo & 1  & 0  \\
        Citroen & 1  & 1  \\
        BMW   & 2  & 2  \\
        Fiat   & 3  & 0  \\
        Mercedes-Benz & 2 & 2 \\
        Peugeot & 10 & 9 \\
    \bottomrule
\end{tabular}}
\end{table}

Based on these figures, we observe that the success of the requirement check is very indicative of the overall eligibility except for Fiat and Alfa Romeo. Note that we have also tested 2 cars that fulfilled the requirements check except the year of production but none of them were eligible.

\subsection{Design of the Tests}

The tests were designed to assess the consent flow and collect information from cars from multiple OEMs. To include cars from multiple OEMs is important because they do not deliver the same data points. Additionally, we also tried to have multiple models of the same OEMs as there are also some variations among models, this is however mostly related to the production year of the car.

In terms of data collection, we have to define what will be the data that we will record on the platform. Our strategy is to include as many points as possible. For this, we will mainly include the data delivered by the three flows (i.e. consent flow, notification flow and request flow) to collect data. In particular, the request flow will be triggered by the reception of a notification. This request will ask for all data points available at that moment.

Beyond this core principle, we still have to take into account particularities among OEMs. A notable difference concerns Mercedes Benz and the other OEMs (Audi, BMW, Ford, Mini, Porsche and Stellantis group). Mercedes Benz only gives access to the odometer (through a request) and it might be done only twice a day. Other OEMs give access to more data points. This includes the GPS coordinates for all of them. The limit is up to 50 request per minute. See table \ref{tab:data-points} for more information concerning data points provided by BMW and Mercedes Benz. Note that the particularities of Mercedes Benz is a choice and not a technical constraint as more data points, including GPS coordinates, are available for Mercedes Benz fleet vehicles. Following our discussion with experts, this choice is related to privacy issues.

Following our strategy of collecting as many data points as possible, we adjust the data collection according to the constraints specified above. For Mercedes Benz, we will collect the odometer twice a day. We choose to collect it early in the morning (5 a.m.) and in the beginning of the night (10 p.m.). This choice is based on our literature review : this data collection make it possible to compute the number of kilometers driven during the night, which is recognised as a significant risk factor \cite{baecke2017,huang2019}. For BMW, we trigger a request to collect data points when receiving a notification. This is most of the time the change of location of the vehicle.

\subsection{Test Results}

\subsubsection{Consent Flow}

The test flow has been tested for 2 BMWs and 2 Mercedes Benz. Drivers were recruited among the participants of the research project. They have reported that these flows are easy to do and quite fast. In both cases, the driver has to confirm his/her consent on the manufacturer platform (i.e. Mercedes Me or BMW Connected Drive). It takes a couple of minutes to do it. For cars of the Stellantis group, we did not have the opportunity to test it but we have received the documentation of the consent flow by the third-party. 

Compared to the cases of BMW and Mercedes, the flow is much more complicated. On the third-party platform, the driver has to confirm his/her identify by using a dedicated website (i.e. \copyright Electronic Identification). Then, this person has to configure the privacy settings in the vehicle to permit the transmission of data if it has not already been done previously for other services. The way to do it depends on the mechanism installed in the vehicle namely double push (need to press at the same time on assistance and SOS button to make the configuration) or screen oriented (3 possibilities according to the interface of the vehicle). Note that, based on the VIN, the insurance company can check if a change in the privacy settings is necessary and, if necessary, inform the driver which of the two mechanisms is installed in the vehicle. Next, the driver has to test the transmission of the data of the vehicle to the telematics server of the OEM. For this, the driver has to push the assistance button to get a call with an operator. According to the documentation provided, this steps should last six minutes from the start-up to the shutdown of the engine. If it does not work, the driver has to go to a Stellantis workshop to fix it. Finally, after a background process of several days that also requires that the car is driven, the driver receives an email to report the mileage of the vehicle to start the connection with the platform of the insurance company. Note that this report of the odometer has to be done again every three months to keep the connection with the service.

\subsubsection{Collection of Data}

We had the opportunity to collect the data of 4 cars. Even if this number is quite low, this collection is still valuable to highlight the diversity of situations across OEMs. Indeed, as discussed above, BMW is representative of most OEMs present in this market of connected cars. These OEMs offer a wide variety of usage information on vehicles, while Mercedes Benz only offers to retrieve the odometer. Data collection is done from enrollment time of the driver till 2022-06-30. The statistics of data collection are displayed in Table~\ref{tab:stats-data}.

\begin{table}
\centering%
\caption{Statistics of data collection by car}\label{tab:stats-data}
\resizebox{0.45\textwidth}{!}{%
\begin{tabular}{c M{0.1\textwidth} c}
    \toprule%
        {\centering\bfseries Car} & {\centering\bfseries Collection Time (days)} & {\centering\bfseries Data Points}\\
    \midrule
        BMW X5 & 63 & 512\\
        BMW 116d & 164 & 1141\\
        Mercedes GLA   & 63 & 76 \\
        Mercedes GLE   & 34 & 26 \\
    \bottomrule
\end{tabular}}
\end{table}

\begin{table}
\centering%
\caption{Data points available by brand. Data points collected during test period appear in italic}\label{tab:data-points}
\resizebox{0.45\textwidth}{!}{%
\begin{tabular}{M{0.05\textwidth} M{0.15\textwidth} M{0.15\textwidth}}
    \toprule%
        {\centering\bfseries Brand} & {\centering\bfseries Data points\\(by notification)} & {\centering\bfseries Data points\\(on request)}\\
    \midrule
        Mercedes & revoke of consent & \textit{odometer}\\
    \midrule
        BMW & accident reported, battery warning, breakdown reported, emergency reported, engine changed, maintenance changed, revoke of consent, \textit{location change} & outside temperature, brake~fluid~change~date, acceleration~evaluation, driving~style, \textit{doors~lock~state, hood~position, GPS~coordinates, heading, odometer, distance~to~next~maintenance, fuel volume}\\
    \bottomrule
\end{tabular}}
\end{table}

\subsubsection{Use Case: Insurance Product}

\begin{figure}[ht!]\vspace*{4pt}
\centerline{\includegraphics[width=70mm,scale=0.5]{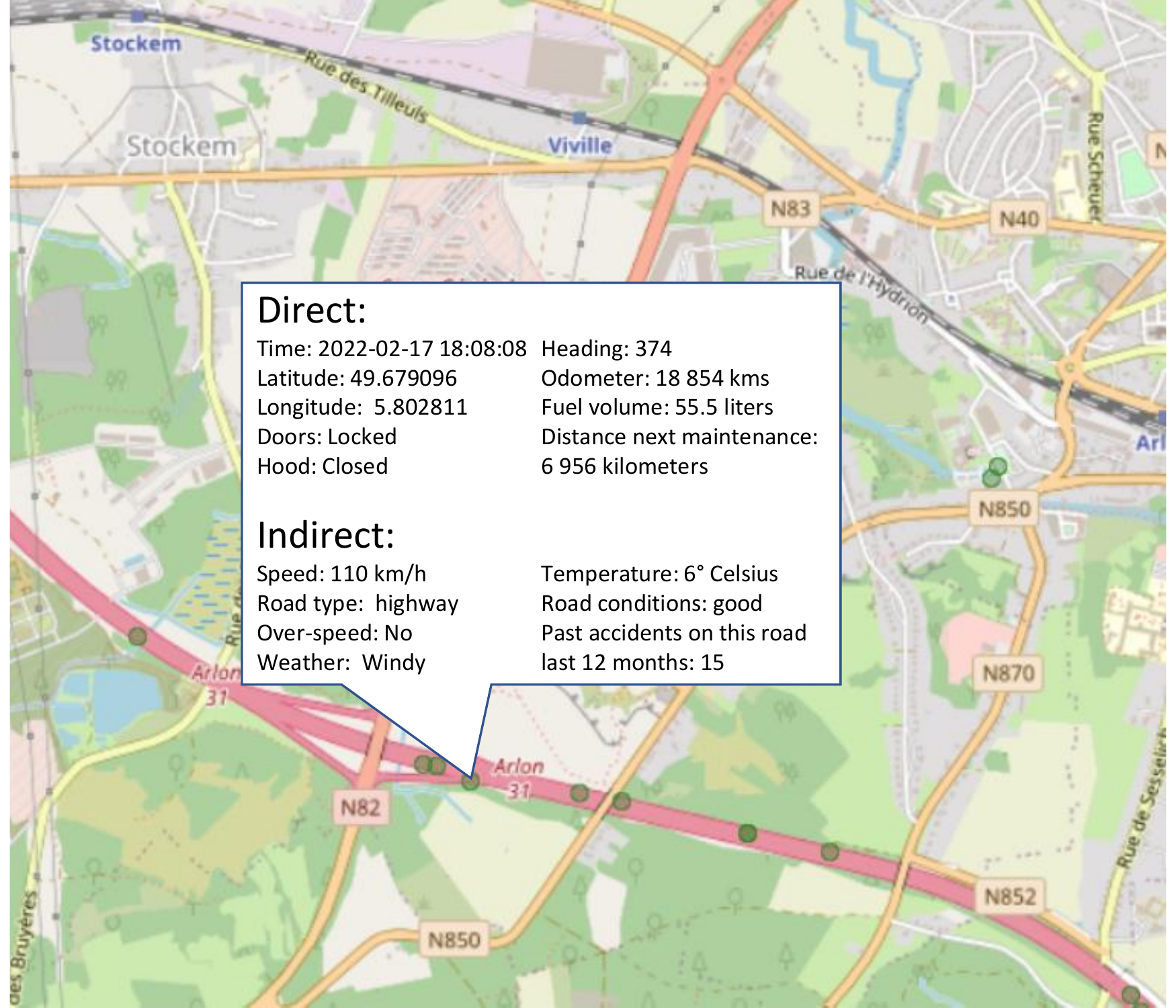}}
\caption{\label{fig:trajectories}Focus on a coordinate during a traject of BMW 116d with illustrative data points}\vspace*{-6pt}
\end{figure}

To highlight the potential of connected cars for insurance products, Figure~\ref{fig:trajectories} shows one the trajectories of a BMW during a day. This figure highlights that rich information can be retrieved from the connected car during its journey for this purpose. Some of them can be defined only based on the data collected. It includes the total distance travelled and the distance travelled by moments of the day (notably during the day and the night). The speed and harsh braking events could be also assessed with a good precision if there are a sufficient number of requests. This information also includes warnings that an accident probably happened trough notifications (i.e. accident, breakdown or emergency reported). Accidents of low severity are typically not reported to the insurance company to avoid an increase of premium \cite{boucher2009} but it is indicative of the driver driving behavior. This is even more the case that car insurance policies usually include a deductible on the reimbursement. Note that BMW also proposes acceleration evaluation and driving style data points but these features were not tested as this is specific to only one OEM.

Beyond these features directly captured by the car, more insights can be retrieved through data augmentation \cite{european2020}. For instance, based on the GPS coordinates, over-speed can be deduced by comparing the speed to the road speed limit as well as the type of road driven. When adding timestamp to the GPS coordinates, weather and traffic conditions can also be defined. Finally, these additional data can also be used to get more data. For example, statistics of past accidents can be retrieved for the roads driven.

Finally, besides giving input to car accident models, the information retrieved might be also valuable for others tasks linked to car insurance products. For instance, it might be used for investigation related to stolen vehicles. Basically, if criminals do not disconnect the car, tracking the car will be easy. If it is not the case, the insurance company will still have more information to make investigations. They might for example analyse to which extend it might be fraud committed by the driver by checking the last lock status of the car or by analysing the last trajectory available. 

\section{DISCUSSION}\label{sec:discussion}

Based on the test of the platform and pricing information received, we can now further compare this new embedded system with existing technologies. We do it for the factors discussed in the related works namely cost, reliability of the system and customer engagement. With respect to the cost, this technology does not have any supplementary monetary fixed cost as the device is already in the car. There is however an usage cost which has to be paid to the OEM and the third-party (if a third-party is chosen). The current rate is around 6.5 euros per month for OEMs such as BMW which gives access to many data points. This cost is quite high for the case of insurance policies. In Grand Duchy of Luxembourg for instance, it represents approximately 8\% of the insurance premium \cite{expatica2021}. In the short run, this price is only economically viable for luxurious cars (leading to higher premiums), especially if theft is included in the car insurance. In the long run, this rate might decrease when the market is more mature. Note that Mercedes Benz proposes a cheaper alternative. The cost is 2.1 euros but it is only possible to develop a PAYD car insurance based on the number of kilometers driven along with a focus on the number of kilometers driven during the night.

In terms of reliability, this technology allows to get rich insights on the way the vehicle is driven. This is even more the case if data directly collected from the car is augmented with external data \cite{european2020}. There are differences in terms of data points collected which makes it difficult to develop a single use case. Furthermore, from a technical point of view, it is advised to use the services of a third-party to retrieve data from the different OEMs. They have their own infrastructure and API specifications which makes it difficult to directly connect a platform to their data flow. During the data collection, we experienced a technical issue for Mercedes Benz that could not be solved easily. In a production environment where insurance premiums would be computed based on usage metrics, contracts for data delivery services with the third-party should include a service level agreement that ensure the continuity of services of the whole process occurring between the cars and the insurance platform.

Finally, concerning customer engagement, this system is currently a doubled-edged sword. On the one hand, this technology has two advantages. First, no additional device has to be installed which might contribute to reduce the perception of the driver to be under scrutiny. Second, this system is much more transparent. The driver participates to the enrollment and he/she has the opportunity to review the data points that will be provided. Furthermore, this enrollment can be revoked any time. On the other hand, the enrollment itself might be cumbersome for the driver depending on the OEM. For Stellantis vehicles, this is currently a blocking point. Even in the presence of very motivated drivers, it would probably require significant dedicated staff from the insurance to help them in this process, leading to additional costs.

\section{CONCLUSION}\label{sec:conclusion}

The recent EU regulation around “eCall” system has given rise to the opportunity to collect data from recent cars. In this paper, we instigated to which extent this technology is mature for usage-based car insurance. To this end, we have tested the eligibility of 19 recent cars, developed a platform to collect the data of 4 connected cars and we have also acquired specific relevant documentation.

Our results highlight that even though this technology is very promising in the future to collect rich data about the usage of the car, it is not yet mature enough to deploy it on a large scale for usage-based car insurance. This result is mainly due to the price, the lack of uniformity of data collected and the enrollment process for which there is still room for improvement. We also highlight that using a third-party data aggregator is probably a good choice to translate the evolving technical choices of each OEM into a more stable single data collection flow. From a managerial perspective, we suggest to insurance companies to keep an eye on the evolution of data points prices and enrollment processes to decide when to enter on the UBI market using this technology. That being said, the technology could still be leveraged today for high value use cases such as insurance policies for luxurious cars, especially if theft is included. This would however require an appropriate service level agreement to ensure the stability of services. 

In terms of limitation, the number of cars connected to our platform is limited (only 4 cars) leading to generalisation issues. We however believe that this issue is mitigated by the fact that we were able to test the eligibility for more cars (i.e. 19 cars) and that these cars are representative of the diversity of situations across OEMs in terms of data points available. Future works might further investigate the customer perspective i.e. customer engagement when using this new technology once the enrollment process has gained more maturity. This could be done through large scale tests of enrollment.

\section*{ACKNOWLEDGMENT}

This work is supported by the Fonds National de la Recherche, Luxembourg (BRIDGES19/IS/14290833). The authors would additionally like to thank Foyer Assurances Luxembourg for their financial support.

\bibliographystyle{vancouver}
\bibliography{sample}

\begin{thebibliography}{10}

\bibitem{haberle2015}
H{\"a}berle T, Charissis L, Fehling C, Nahm J, Leymann F.
\newblock The connected car in the cloud: a platform for prototyping telematics
  services.
\newblock IEEE Software. 2015;32(6):11-7.

\bibitem{gao2021}
Gao C, Wang G, Shi W, Wang Z, Chen Y.
\newblock Autonomous Driving Security: State of the Art and Challenges.
\newblock IEEE Internet of Things Journal. 2021.

\bibitem{baecke2017}
Baecke P, Bocca L.
\newblock The value of vehicle telematics data in insurance risk selection
  processes.
\newblock Decision Support Systems. 2017;98:69-79.

\bibitem{denuit2007}
Denuit M, Mar{\'e}chal X, Pitrebois S, Walhin JF.
\newblock Actuarial modelling of claim counts: Risk classification, credibility
  and bonus-malus systems.
\newblock John Wiley \& Sons; 2007.

\bibitem{stigson2014}
Stigson H, Hagberg J, Kullgren A, Krafft M.
\newblock A one year pay-as-you-speed trial with economic incentives for not
  speeding.
\newblock Traffic injury prevention. 2014;15(6):612-8.

\bibitem{edlin1999}
Edlin A.
\newblock Per-mile premiums for auto insurance.
\newblock NBER Working Paper Series. 1999.

\bibitem{boucher2009}
Boucher JP, Denuit M, Guillen M.
\newblock Number of accidents or number of claims? An approach with
  zero-inflated Poisson models for panel data.
\newblock Journal of Risk and Insurance. 2009;76(4):821-46.

\bibitem{denuit2019}
Denuit M, Guillen M, Trufin J.
\newblock Multivariate credibility modelling for usage-based motor insurance
  pricing with behavioural data.
\newblock Annals of Actuarial Science. 2019;13(2):378-99.

\bibitem{huang2019}
Huang Y, Meng S.
\newblock Automobile insurance classification ratemaking based on telematics
  driving data.
\newblock Decision Support Systems. 2019;127:113156.

\bibitem{ullah2020}
Ullah S, Kim DH.
\newblock Lightweight driver behavior identification model with sparse learning
  on in-vehicle can-bus sensor data.
\newblock Sensors. 2020;20(18):5030.

\bibitem{castignani2015}
Castignani G, Derrmann T, Frank R, Engel T.
\newblock Driver behavior profiling using smartphones: A low-cost platform for
  driver monitoring.
\newblock IEEE Intelligent transportation systems magazine. 2015;7(1):91-102.

\bibitem{handel2014}
Handel P, Skog I, Wahlstrom J, Bonawiede F, Welch R, Ohlsson J, et~al.
\newblock Insurance telematics: Opportunities and challenges with the
  smartphone solution.
\newblock IEEE Intelligent Transportation Systems Magazine. 2014;6(4):57-70.

\bibitem{eu2014}
{European Parliament}.
\newblock Decision N° 585/2014/EU of the European Parliament and of the
  Council of 15 May 2014 on the deployment of the interoperable EU-wide eCall
  service (Text with EEA relevance).
\newblock Official Journal of the European Union. 2014.

\bibitem{mcdonnell2021}
McDonnell K, Murphy F, Sheehan B, Masello L, Castignani G, Ryan C.
\newblock Regulatory and Technical Constraints: An Overview of the Technical
  Possibilities and Regulatory Limitations of Vehicle Telematic Data.
\newblock Sensors. 2021;21(10):3517.

\bibitem{bian2018}
Bian Y, Yang C, Zhao JL, Liang L.
\newblock Good drivers pay less: A study of usage-based vehicle insurance
  models.
\newblock Transportation research part A: policy and practice. 2018;107:20-34.

\bibitem{paefgen2013}
Paefgen J, Staake T, Thiesse F.
\newblock Evaluation and aggregation of pay-as-you-drive insurance rate
  factors: A classification analysis approach.
\newblock Decision Support Systems. 2013;56:192-201.

\bibitem{guillen2019}
Guillen M, Nielsen JP, Ayuso M, P{\'e}rez-Mar{\'\i}n AM.
\newblock The use of telematics devices to improve automobile insurance rates.
\newblock Risk analysis. 2019;39(3):662-72.

\bibitem{tselentis2016}
Tselentis DI, Yannis G, Vlahogianni EI.
\newblock Innovative insurance schemes: pay as/how you drive.
\newblock Transportation Research Procedia. 2016;14:362-71.

\bibitem{litman2007}
Litman T.
\newblock Distance-based vehicle insurance feasibility, costs and benefits.
\newblock Victoria. 2007;11.

\bibitem{kaiser2018}
Kaiser C, Stocker A, Festl A, Lechner G, Fellmann M.
\newblock A research agenda for vehicle information systems.
\newblock Twenty-Sixth European Conference on Information Systems (ECIS2018).
  2018.

\bibitem{eu}
The European~Parliament CoEU.
\newblock On the protection of natural persons about the processing of personal
  data and on the free movement of such data, and repealing Directive 95/46/EC
  (General Data Protection Regulation).
\newblock Official Journal of the European Union. 2016;679:1-88.

\bibitem{european2020}
Board EDP. Guidelines 1/2020 on Processing Personal Data in the Context of
  Connected Vehicles and Mobility Related Applications. Version adopted for
  public consultation Brussels; 2020.

\bibitem{hardt2012}
Dick H. The OAuth 2.0 Authorization Framework; 2012.
\newblock \url{https://datatracker.ietf.org/doc/html/rfc6749}.

\bibitem{expatica2021}
Gary B. Car insurance in Luxembourg; 2021.
\newblock
  \url{https://www.expatica.com/lu/finance/insurance/car-insurance-luxembourg-82882/}.

\end{thebibliography}
\end{document}